\documentclass[aps,pra,twocolumn,showpacs, amsmath, amssymb]{revtex4}

\usepackage{graphicx}
\usepackage{dcolumn}
\usepackage{bm}
\usepackage[utf8]{inputenc}
\usepackage{siunitx}
\usepackage{braket}
\usepackage{breqn}
\usepackage{amsmath}
\usepackage{wrapfig}
\usepackage{xcolor}
\usepackage[version=3]{mhchem}
\usepackage{bbold}
\usepackage{mathtools}
\usepackage{tikz}
\usepackage{float}
\usetikzlibrary{arrows}

\newcommand\SmallMatrix[1]{{%
  \small\arraycolsep=0.6\arraycolsep\ensuremath{\begin{pmatrix}#1\end{pmatrix}}}}

\graphicspath{{pix/}}

\makeatletter
\let\cat@comma@active\@empty
\makeatother

\begin{document}


\title{Generalized Effective Operator Formalism for Decaying Systems}

\author{Marius Paraschiv}
 \author{Sabine W{\"o}lk}
\author{Thomas Mannel}
\author{Otfried G{\"u}hne}
\affiliation{%
Naturwissenschaftlich-Technische Fakult{\"a}t, Universit{\"a}t Siegen, Walter-Flex-Str. 3, 57068 Siegen, Germany
}%
\date{\today}

\begin{abstract}
\noindent
Systems of neutral kaons can be used to observe entanglement and the violation
of Bell inequalities. The decay of these particles poses some problems, however,
and recently an effective formalism for treating such systems has been derived.
We generalize this formalism and make it applicable to other quantum systems
that can be made to behave in a similar manner. As examples, we discuss  two
possible implementations of the generalized formalism using trapped ions such
as $^{171}$Yb or $^{172}$Yb, which may be used to simulate kaonic behavior in
a quantum optical system.
\end{abstract}


\maketitle


\section{Introduction}
\noindent
The unexpected effects of quantum correlations were initially described by
Einstein, Podolsky and Rosen in their 1935 paper \cite{einstein}
as a phenomenon questioning the completeness of quantum mechanics.
The natural proposal to overcome the problems raised therein was
to assume the existence of hidden variables: additional parameters
are not present in the quantum mechanical description, but which
characterize the behavior and evolution of a quantum system.

In 1964, however, John Bell \cite{bell} proposed an inequality that
must be satisfied by any local hidden-variable alternative to quantum
mechanics. It turns out that quantum mechanics violates this inequality,
and thus, no local hidden-variable theory can reproduce its full
predictions. Bell's inequality was further generalized in 1969 by
Clauser, Horne, Shimony, and Holt \cite{chsh}, and many versions of
Bell-type inequalities have been proposed since then \cite{sliwa, collins,
variousbell}.


EPR-correlations have been a research topic in the field of particle physics. In particular,
the system of neutral kaons exhibits interesting quantum correlations 
which have been studied since the discovery of the kaons in the 1960's.   
Due to the strangeness quantum number the neutral kaon and ist antiparticle
are different states, allowing for a description as a  two state system. However, 
due to the decay of the kaons, the time-evolution of the two-state system is 
non-unitary. The details can be found in particle physics textbooks 
(see eg. Ref.~\cite{Nachtmann}); in the context of quantum optics this has been 
discussed in  Ref.~\cite{bertlmann}.

Treating the system of neutral kaons as a two-state system allows 
us to draw some analogies to photons \cite{gisin, go}, but existing approaches 
suffered from a series of
shortcomings \citep{bertlmann2}. The most important issues were problems
with the normalization of the state of the decaying system, the difficulty
of choosing an active method to measure the quasi-spin \cite{bramon}
and the difficult  problem of a generalization to a higher number of
particles.

In Ref.~\cite{didomenico} the authors proposed a reformulation of the
problem in terms of an effective operator formalism. Here, the time evolution
of the system as well as the measurement angles can be incorporated into an
effective operator that offers a series of advantages over the direct photon-analogy
method: By including the non-unitary time evolution into the effective operator,
one ensures that normalization is only performed with respect to the surviving
particles. Another advantage is the possibility to generalize it to an arbitrary
number of particles just by the usual tensor product. Finally, an interesting
property of neutral kaons, namely the violation of CP-symmetry is easily
included within the formalism.

In this paper, we present a generalization of this approach to general decay
processes and apply it to trapped ion systems. In detail, the paper is organized
as follows:  In Section II we describe the physics of neutral kaons and the
existing approaches to effective operators. In Section III we describe our
generalized approach. Section IV presents an application to kaons and various
Bell inequalities in this setting. Section  V describes applications to
trapped ions such as $^{171}$Yb or $^{172}$Yb, which may be used to simulate
kaonic behavior in quantum optics. Finally, we conclude and discuss possible
further directions of research.

\section{Entanglement of neutral kaons}
\subsection{Neutral kaons}

In this section we give a brief summary of the quantum mechanics of 
the neutral kaon system, which sets the stage for the further considerations. 
The neutral kaons are composed of a strange quark and a down anti-quark, 
where the strange quark carries a quantum number called strangeness.  Due 
this quantum number we can distinguish the neutral kaon from its antiparticle  
\begin{eqnarray}
&& \Ket{K^0} = \Ket{\bar{d} s}\, , \quad  S\Ket {K^0} = +\Ket{K^0},\\
&& \Ket {\bar{K}^0}  = \Ket{\bar{s} d} \, , \quad S\Ket {\bar{K}^0} = -\Ket{\bar{K}^0} \, .
\end{eqnarray}
The second relevant quantum number is related to the behaviour of the kaons 
under charge conjugation $C$ and parity $P$. Since the kaons are pseudoscalar 
particles, one obtains (making a choice for a possible arbitrary phase) 
\begin{eqnarray}
&& CP\Ket {K^0} = -\Ket{\bar{K}^0},\\ \label{cpop1}
&& CP\Ket {\bar{K}^0} = -\Ket{K^0}. 
\end{eqnarray} 
In particular, $S$ and $CP$ do not commute and hence there are no common 
eigenstates; in fact, we may construct the CP eigenstates from the strangeness 
eigenvectors to be 
\begin{eqnarray} 
\Ket {K_1^0} = \frac{1}{\sqrt{2}}(\Ket {K^0} - \Ket {\bar{K}^0}), \label{CP1}  \label{CP1}\\
\Ket {K_2^0} = \frac{1}{\sqrt{2}}(\Ket {K^0} + \Ket {\bar{K}^0})  \label{CP2}. \label{CP2}
\end{eqnarray} 

Neutral kaons are produced by generating an $s \bar{s}$ pair, which 
hadronizes  into strange particles. A particularly clean way is pursued at 
$DA\Phi NE$ at Frascati: The kaons are generated from the decay of a
$\phi$ meson, which consists of two strange quarks, but is heavy enough to 
(exclusively) decay into a pair of neutral kaons. The $\phi$ meson has a definite $CP$ quantum number and the kaon state in 
the moment of the decay  of the $\phi$ ($t=0$) is given by  
\begin{equation}\label{eq:1}
\Ket{\psi(t=0)} = \frac{1}{\sqrt{2}} (\Ket{K^0}\Ket{\bar{K}^0} - \Ket{\bar{K}^0}\Ket{K^0}).
\end{equation}


Kaons decay through weak interaction processes, the relevant weak transition is
the decay of the strange quark into an up quark. Hence the possible final states are either  
two or three pions, since there is no phase space for heavier states. However, the two- 
and three-pion states with vanishing orbital angular momentum $\ell = 0$ have 
different CP eigenvalues, 
\begin{eqnarray} 
&& CP \Ket{\pi \pi (\ell = 0)} = \Ket{\pi \pi (\ell = 0)},  \\
&& CP \Ket{\pi \pi \pi (\ell = 0)} = - \Ket{\pi \pi \pi (\ell = 0)}. 
\end{eqnarray} 
Thus, if $CP$ were conserved, the $K_1^0$ could decay exclusively to two pions, 
while  $K_2^0$ could only decay into three pions. Since the phase space for the decay 
into three pions is much smaller than the one for the decay into two pions, the $K_2^0$ 
has a significantly longer lifetime.   

Weak interactions mediate not only the decays of the kaons, but also an effect 
called mixing. Since the $K^0$ and the $\overline{K}^0$ have common decay channels $f$, 
a  $K^0$ state can oscillate into a $\overline{K}^0$ state through the process
$K^0 \to f \to \overline{K}^0$.  Thus the neutral kaon states undergo a mixing which in the 
two-dimensional space of $K^0$ and $\overline{K}^0$ is described by a hamiltonian
\begin{equation}\label{eq:77}
H = M + i \Gamma,
\end{equation}
where $M$ and $\Gamma$ are hermitean $2 \times 2$ matrices. Note that this hamiltonian 
takes into account the decay of the kaons through the contribution of $\Gamma$, which makes 
the total hamiltonian non-hermitean. 

Assuming first that CP is a good quantum number, the two $CP$ eigenstates (\ref{CP1}) and 
(\ref{CP2}) are the eigenstates of the hamiltonian, and the eigenvalue equation reads
\begin{equation}\label{eq:8}
H\Ket{K_{S,L}} = \lambda_{S,L}\Ket{K_{S,L}}.
\end{equation}
The eigenvalues are complex, since $H$ is non-hermitean. We decompose them into 
real and imaginary parts as   
\begin{equation}
\lambda_{S,L} = m_{S,L} - \frac{i}{2}\Gamma_{S,L},
\end{equation}
where $m_{S,L}$ are the masses of the short and long-lived states and
$\Gamma_{S,L} \ge 0$ are the decay widths. Note that $\Gamma_S < \Gamma_L$, so 
$\Gamma_S$ is the width of the short-lived, while $\Gamma_L$ is the one of the 
long-lived  kaon. Furthermore, if CP were conserved, we would have 
$\Ket{K_S} =  \Ket {K_1^0} $  and $\Ket{K_L} =  \Ket {K_2^0}$. 

The eigenvalue problem (\ref{eq:8}) is non-hermitean, which not only leads to complex 
eigenvalues, but also the eigenvectors are not orthogonal to each other. We make a choice 
of the relative phases of $K_L$ and $K_S$ as 
\begin{eqnarray}
&& \langle K_S | K_S \rangle = \langle K_L | K_L \rangle = 1 \\ 
&& \langle K_S | K_L \rangle = \langle K_S | K_L \rangle^*  \ge 0  
\end{eqnarray}  

Weak interactions violate the CP symmetry; in fact this was discovered in the system of 
neutral kaons. As a consequence, $CP$ and $H$ do not commute and thus the CP eigenstates 
(\ref{CP1},\ref{CP2}) are not identical to the eigenstates $K_L$ and $K_S$ of the hamiltonian.
To this end, the eigenstates of the hamiltonian become 
\begin{equation}\label{eq:5}
\begin{split}
 \Ket {K_S} = \frac{1}{N}(p \Ket {K^0} - q \Ket {\bar{K}^0}), \\
 \Ket {K_L} = \frac{1}{N}(p \Ket {K^0} + q \Ket {\bar{K}^0}).
\end{split}
\end{equation}
with $N = \sqrt{|p|^2+|q|^2}$,  and only if CP is conserved we have $p=q$.   
However, CP violation is a small effect; rewriting the eigenstates of the hamiltonian 
$K_S$ and $K_L$ in terms of the CP eigenstates $K_1^0$ and $K_2^0$ 
\begin{equation}\label{eq:kskl}
\begin{split}
\Ket{K_S} &= \frac{1}{\sqrt{1+|\epsilon|^2}} \left(\Ket{K^0_1} + \epsilon \Ket{K^0_2}  \right)   \\ 
\Ket{K_L} &= \frac{1}{\sqrt{1+|\epsilon|^2}} \left(\Ket{K^0_2} + \epsilon \Ket{K^0_1}  \right) 
\end{split}
\end{equation} 
we define the $CP$ violating parameter $\epsilon$, which has a value of approximately $\epsilon
\approx 10^{-3}$. The fact that there is CP-violation observed in the neutral kaon system,
shown here by the non-zero $ \epsilon $ parameter leads to a slight non-orthogonality of
the short and long-lived states.

Including $CP$ violation, the  total time evolution of the kaon system is then
given by
\begin{equation}\label{eq:10}
\begin{split}
 \Ket {K^0(t)} = g_+(t) \Ket {K^0} + \frac{q}{p} g_-(t)\Ket {\bar{K}^0}), \\
 \Ket {\bar{K}^0(t)} = \frac{p}{q} g_-(t) \Ket {K^0} + g_+(t) \Ket {\bar{K}^0}).
\end{split}
\end{equation}
with the time-dependent functions
\begin{equation}\label{eq:11}
\begin{split}
 g_+(t) &= \frac{1}{2}(e^{-i\lambda_St}+e^{-i\lambda_Lt}), \\
 g_-(t) &= \frac{1}{2}(-e^{-i\lambda_St}+e^{-i\lambda_Lt}).
\end{split}
\end{equation}
The lifetime difference in the kaon system is substantial: The lifetimes of
the two states are: $\tau_S = 8.95 \times 10^{-11}s$ for the short-lived state
$\Ket{K_S}$ and $\tau_L = 5.11 \times 10^{-8}s$ for the long-lived state $\Ket{K_L}$.

There are two physical properties of neutral kaons that make them important for
our model: the first one is decay as the system is represented by the two states
with different lifetimes. The second property is flavour oscillation. As the
state $\Ket{\psi}$ of Eq.~(\ref{eq:1}) evolves in time, the particles undergo
mixing, essentially an oscillation between particle and antiparticle
(See Eq.~(\ref{eq:10})).
This
type of behaviour is known as neutral particle oscillation and will prove
useful to us, as we will want to make our measurements at different times,
by performing the measurement on one particle and allowing the other to
evolve for an additional time $\tau$, before measuring. This effectively allows
to measure the spin in different directions.

\subsection{Brief description of the effective formalism}

For a system of two particles of spin-1/2, one can select four settings
for the directions of the corresponding spins (two for each particle,
say $A_1$ and $A_2$, for Alice's particle and $B_1$ and $B_2$, for
Bob's particle). These can be used as parameters for the CHSH inequality
\cite{chsh}
\begin{equation}
\langle A_1B_1\rangle + \langle A_2B_1\rangle + \langle A_1B_2\rangle - \langle A_2B_2\rangle \leq 2,
\end{equation}
where the bound is valid for local realistic theories and can be
violated by quantum mechanics.

One can define a quasi-spin quantity, for neutral kaons, with the parametrization
\begin{equation}
\Ket {k_{n}} = \cos(\frac{\alpha_{n}}{2}) \Ket {K_{S}}+ \sin(\frac{\alpha_{n}}{2}) e^{i\phi_{n}} \Ket {K_{L}},
\end{equation}
and thus give the possibility to the experimenter of choosing different quasi-spin directions. One must be careful though, there is no arbitrary spin direction in this case, one only has the choice of $\Ket {K_S}$, $\Ket {K_L}$, $\Ket {K^0}$, $\Ket {\bar{K}^0}$, due to the fact that only strangeness or lifetime measurements can be performed.

For a certain quasi-spin direction $\Ket {k_n} $ at a certain measurement time $t_n$, the expectation value can be written in terms of the probability of obtaining a $\Ket {k_n}$ [denoted here by Y(yes) $|$ N(no)] as
\begin{equation}\label{eq:14}
\begin{split}
E(k_{n},t_{n}) & = P(Y:k_{n},t_{n}) - P(N:k_{n},t_{n}) \\
& = 2P(Y:k_{n},t_{n})-1.
\end{split}
\end{equation}
Noteworthy is the fact that the probability $ P(N:k_{n},t_{n}) $ does not only include the case of detecting a $\Ket{k_n}$, but also non-detection events.

Using our parametrization and the fact that
\begin{equation}\label{eq:15}
P(Y:k_{n},t_{n}) = Tr[\Ket {k_{n}} \Bra {k_{n}} \rho(t_n)],
\end{equation}
we can find an operator that satisfies the property
\begin{equation}\label{eq:16}
E(k_n,t_n) = Tr[O^{\rm eff}(k_n,t_n)\rho(t=0)],
\end{equation}
where $O^{\rm eff}(k_n,t_n)$ is called an "effective operator".

\renewcommand{\thefootnote}{\fnsymbol{footnote}}

With this, the matrix form of the effective operator in the lifetime eigenbasis is 

\begin{align}
& O^{\rm eff} =
\\
\nonumber
&
\SmallMatrix{
{\cos^{2} (\frac{\alpha_{n}}{2}) e^{- \Gamma_S t_n}-1}
&
{\frac{1}{2}\sin(\alpha_n) e^{i(\phi_n - \omega t_n)} e^{- \Gamma t_n}}
\\
{\frac{1}{2}\sin(\alpha_n) e^{-i(\phi_n - \omega t_n)} e^{- \Gamma t_n}}
&
{\sin^{2} (\frac{\alpha_{n}}{2}) e^{- \Gamma_L t_n}-1}},
\end{align}
as can be constructed by following the derivations in Ref.~\cite{didomenico}, with $\omega \propto \Delta m$.
Here, $\Gamma = \frac{1}{2}(\Gamma_S + \Gamma_L)$, where
$\Gamma_{S,L}$ are the decay constants of the short and
long-lived states.

A few remarks are in order: First, from the matrix form of this operator,
we see that for large  times, when the probability that the particles
have decayed is very high, the effective operator tends to minus identity.
This is a reasonable expectation, since, as the particles decay,
the overall amount of detection events decreases. Second, using Eq.~(\ref{eq:16}) it
is easy to see that in the case of a multi-particle system the generalization
is straightforward: $ E = Tr(O_1^{\rm eff} \otimes O_2^{\rm eff}\otimes
... \otimes O_n^{\rm eff}  \rho_n)$.

The experimental set-up of a Bell-test is rather standard: a pair of particles
produced at a source propagate in opposite directions. They are detected by
two experimenters, by tradition named Alice and Bob. There are two possible
ways to test any given Bell-type inequality in this situation:
\begin{itemize}
\item Fix the measurement times and measure for different quasi-spins.
\item Fix the quasi-spins and measure at different times.
\end{itemize}

Because of the scarcity of directions when it comes to choosing a
quasi-spin, the former option does not provide interesting
information, other than what would be, much easier, obtained
with non-decaying systems like photons. However, the decay
and strangeness oscillation make the latter much more
appealing. In short, both Alice and Bob agree upon a
single measurement direction, say $ \bar{K}^0 $ and each
measures his or her particle at a different time.

The measurement procedure in the case of kaons, also
requires a certain amount of discussion: while it
surely is possible to allow the particles to
decay, and then identify the initial particle by
it's decay products (passive measurement) this does
not allow the experimenter a free choice of measurement
times. One possibility (active measurements) would be
to insert a piece of matter in the way of the kaon
beam, and thus, force the particle to decay, by
interaction. The distance between the object and
the kaon source would help set the measurement times.
Another essential aspect is the ability of the
experimenter to choose their measurement angle
(in our case a choice between the particle and
its antiparticle). A more detailed analysis of
the various types of measurements that can be
performed on kaons, is given in Ref.~\cite{bramon}.

\section{General Decaying Systems}

\begin{figure}[t]
  \centering
    \includegraphics[width=0.30\textwidth]{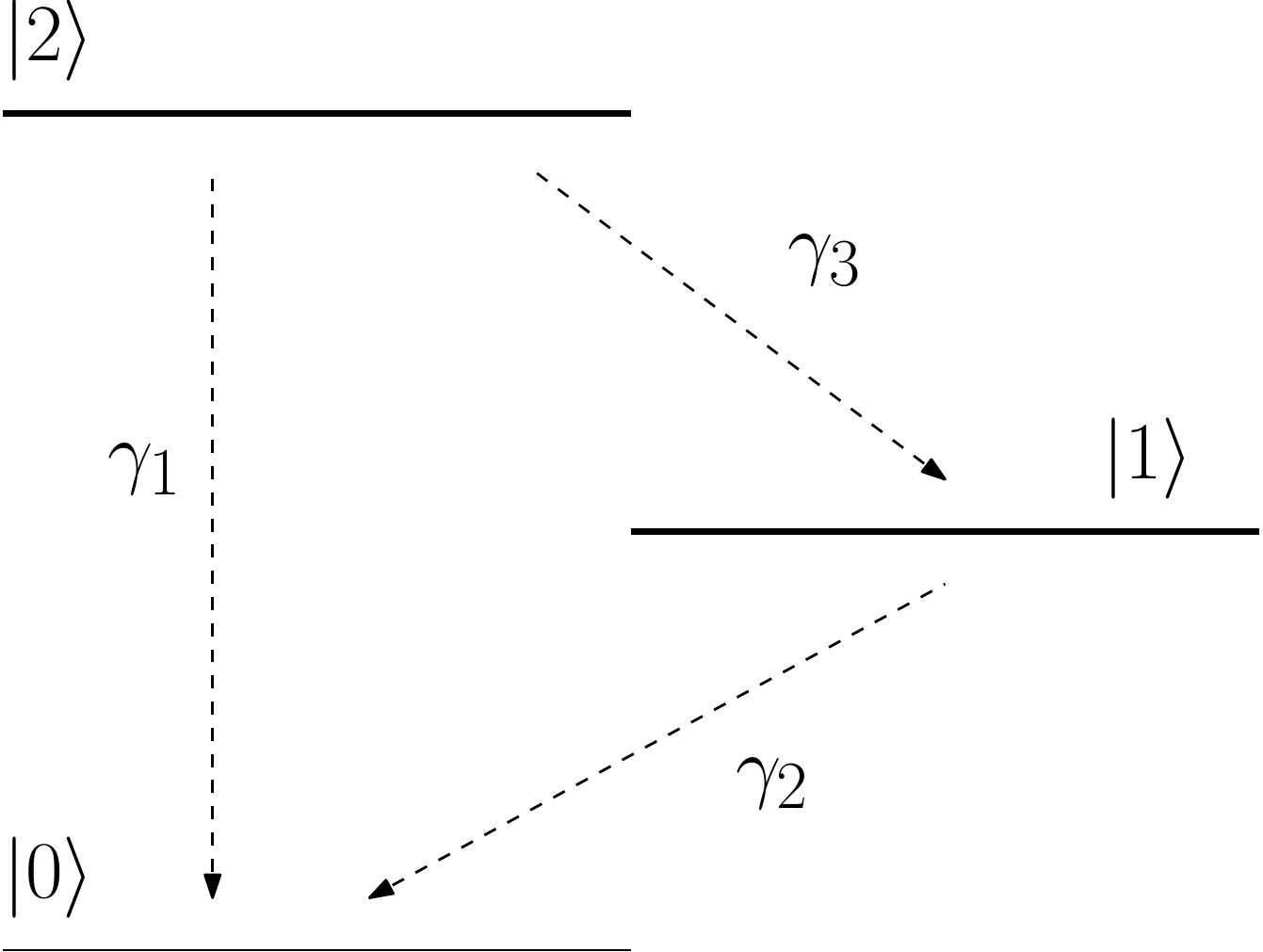}
    \caption{A general three-level system at zero temperature with different decay
    processes. See the text for further details.}
  \label{fig1}
\end{figure}


In this section we describe our generalization of the effective operator formalism. Later, we will
see how it can be applied to other systems beyond neutral kaons. This generalization is based
on the Bloch equation formalism \cite{bloch}. We start by treating a closed three-level
system at zero temperature, with population decay from the upper two levels to a ground level
(see Fig.~\ref{fig1})
The long and short-lived states $\Ket{K_S}$ and $\Ket{K_L}$ would correspond to $\Ket{2}$ and $\Ket{1}$, respectively. The $\Ket{0}$ level plays the role of a general "decayed" level.

The time evolution of a single particle is given by the Lindblad equation
\begin{equation}
\dot{\rho}(t) = -i [H, \rho(t)] - \sum_i {\gamma_i}(\frac{1}{2}\{\Lambda_i^{\dagger}\Lambda_i, \rho(t)\} - \Lambda_i \rho(t) \Lambda_i^{\dagger}),
\end{equation}
with $\Lambda_i$ being jump operators between different levels. Here, $H$ is the mass term $M$ from Eq.(\ref{eq:77}), a notation we will maintain throughout the rest of the paper. For our case, these are given by
\begin{equation}
\begin{split}
&\Lambda_{20} = \Ket {0} \Bra {2}, \\
&\Lambda_{21} = \Ket {1} \Bra {2}, \\
&\Lambda_{10} = \Ket {0} \Bra {1}.
\end{split}
\end{equation}
Taking into account only the unitary part of the time evolution
\begin{equation}
 \dot\rho(t) = -i [H, \rho(t)],
\end{equation}
and denoting the density operator in column vector form as
\begin{equation}
\vec{\rho} = (\rho_{1,1},...,\rho_{1,N},\rho_{2,1},...,\rho_{2,N},...,\rho_{N,N})
\end{equation}
 the unitary part becomes
\begin{equation}\label{eq:22}
 -i [H, \rho] = -i (H \otimes  \mathbb{1} -  \mathbb{1} \otimes H^T ) \vec {\rho}.
\end{equation}
As for the non-unitary part of the time evolution, we simplify the notation by writing it as
\begin{equation}
\dot{\rho} = \frac{-\gamma}{2}(\Lambda_{+}\Lambda_{-} \rho + \rho \Lambda_{+}\Lambda_{-} - 2 \Lambda_{-} \rho \Lambda_{+}),
\end{equation}
where $ \Lambda_{+} $ stands for $\Lambda^{\dagger} $ and $ \Lambda_{-} $ stands for $\Lambda$.

Using the transformation from Eq.~(\ref{eq:22}), we have
\begin{equation}
\begin{split}
& \Lambda_{+}\Lambda_{-} \cdot \rho = (\Lambda_{+}\Lambda_{-} \otimes \mathbb{1} )\vec {\rho} , \\
& \rho \cdot \Lambda_{+}\Lambda_{-} = (\mathbb{1} \otimes (\Lambda_{+}\Lambda_{-})^T) \vec {\rho}, \\
\Lambda_- \rho \Lambda_+  = \Lambda_- &\cdot (\rho \Lambda_+) = (\Lambda_- \otimes \mathbb{1}) \cdot (\mathbb{1} \otimes \Lambda_{+}^T) \vec {\rho}.
\end{split}
\end{equation}
The Lindblad equation can now be written in operator form
\begin{equation}
\vec {\dot{\rho}} = A \vec {\rho},
\end{equation}
with the time evolution operator given by the above results
\begin{equation*}
\begin{split}
A = & [-i (H \otimes  \mathbb{1} -  \mathbb{1} \otimes H^T ) \\
 &- \frac{\gamma}{2}(\Lambda_{+}\Lambda_{-} \otimes \mathbb{1} + \mathbb{1} \otimes \Lambda_{+}\Lambda_{-} -2 \Lambda_{-} \otimes \Lambda_{-} )].
\end{split}
\end{equation*}
Note that, while for the specific way in which we have defined the $\Lambda$ operator, the relation $\Lambda_+^T = \Lambda_-$ holds, this is not true in general.
With this, the time evolution equation of the decaying system is simply
\begin{equation}\label{eq:26}
\vec {\rho} (t) = e^{At} \vec {\rho}(0).
\end{equation}
Worth mentioning is that this formulation of the time evolution also makes a numerical approach towards solving the problem possible.

The final step is to construct the general effective operator. From Eq. (\ref{eq:14}) - (\ref{eq:16}) it follows that
\begin{equation}
E = Tr[2 \Ket {k_n} \Bra {k_n} \rho(t) - \rho(0)],
\end{equation}
where we used the fact that $ Tr[\rho(0)] = 1$.
In order to apply the time evolution, we will write everything in
vector notation and then go back to the original matrix notation,
to recover the final form of the effective operator. Denoting the
matrix $\Ket {k_n} \Bra {k_n}$ as $ K $, we have
\begin{equation}
\begin{split}
E &= Tr[2 ( K \otimes \mathbb{1}) \vec \rho(t) - \vec \rho(0)] \\
  &= Tr[2 ( K \otimes \mathbb{1}) e^{At} \vec \rho(0) - \vec \rho(0)],
\end{split}
\end{equation}
see Eq.~(\ref{eq:26}). In vector notation, one should retain the indices
from the matrix notation ($\rho_{i,j}$), in order for the trace  to
make sense, thus, in the above equation, the trace should be understood as
\begin{equation}
Tr[\vec\rho(t)] = \sum_{i} \vec\rho_{ii}(t).
\end{equation}
Finally, we apply the exponential to $\vec K$ and revert to the
original matrix notation, where we replace $\vec K e^{At} $ with $K(t)$
\begin{equation}
\begin{split}
E &= Tr[(2 K(t) - \mathbb{1} )\rho(0)] \\
  &= Tr[O^{\rm eff} \rho(0)],
\end{split}
\end{equation}
from which one can simply identify the general effective operator as
\begin{equation}
O^{\rm eff}(k_n,t_n) = 2K(k_n,t_n) - \mathbb{1}.
\end{equation}
Here, the dependence of $K$ on the time $t_n$ and measurement direction $k_n$ has been explicitly highlighted.

The above effective operator represents measurements performed on a
single particle, and the possible settings are given by the measurement
"angles" of the quasi-spin and the various times. Now, as previously
mentioned, one can test correlations between larger numbers of
particles, by taking the tensor products of the corresponding
effective operators.

\section{Results for Neutral Kaons}

\begin{figure}
  \centering
    \includegraphics[width=0.47\textwidth]{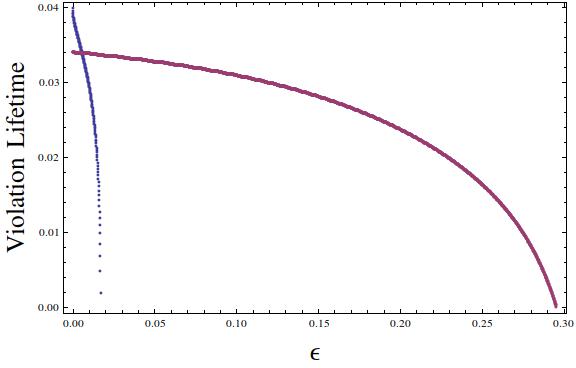}
    \caption{Dependence of the violation lifetime, for the CHSH (fast decaying curve) and SCG (in units of $10$ ns) on the CP-violation parameter.}
  \label{fig2}
\end{figure}

\begin{figure}
  \centering
    \includegraphics[width=0.47\textwidth]{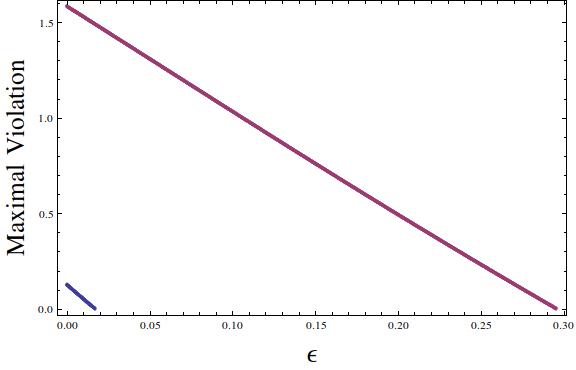}
    \caption{Dependence of the maximal violation of the CHSH (short line) and SCG inequalities on the CP-violation parameter. The measurement settings are the same as for Fig.~\ref{fig2}.}
  \label{fig3}
\end{figure}

\begin{figure}
  \centering
    \includegraphics[width=0.47\textwidth]{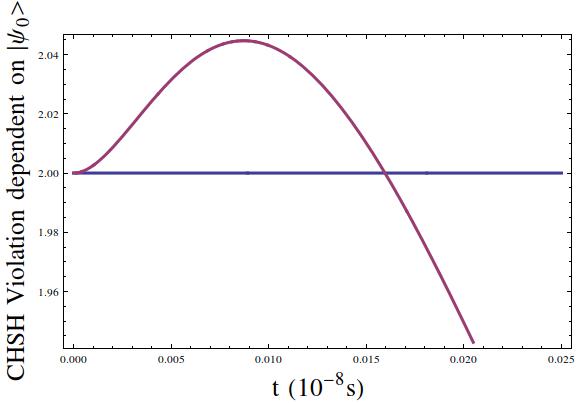}
    \caption{Plot of Eq.~(\ref{chshtrace}) illustrating an obtained violation of the CHSH inequality by starting from an initial state $\Ket{\psi_0}$, described in Eq.(\ref{eq:1}).}
  \label{fig_chsh}
\end{figure}

A first test of the formalism is to apply it for the case of the CHSH inequality on two three-level quantum systems, thus also including the neutral kaon case.

Alice's settings are represented by the indices $A_1$ and $A_2$ while
Bob's settings are represented by $B_1$ and $B_2$. The witness form of the CHSH inequality
is given by
\begin{equation*}
S^{\rm eff} = O^{\rm eff}_{A_1}\otimes (O^{\rm eff}_{B_1}-O^{\rm eff}_{B_2})+O^{\rm eff}_{A_2}\otimes(O^{\rm eff}_{B_1}+O^{\rm eff}_{B_2}),
\end{equation*}
and thus, the test for a possible violation is simply reduced to the consideration of the maximal and minimal eigenvalues of $S^{\rm eff}$, that is, the condition
\begin{equation} \label{chshtrace}
|Tr(S^{\rm eff}\rho)| \leq 2.
\end{equation}

In Fig.~\ref{fig_chsh}, we plot the relation from Eq.~(\ref{chshtrace}), considering the initial state of Eq.(\ref{eq:1}). While a violation is observed (compare with Ref.\cite{gisin}), in order to get the optimal results, we must consider the minimal and maximal eigenvalues of the effective operator, which is the subject for the rest of this section.

Another interesting Bell-type inequality, the Sliwa-Collins-Gisin inequality, was first proposed
by C.~Sliwa \cite{sliwa} and shown by D.~Collins and N.~Gisin \cite{collins} to be non-equivalent
to the original CHSH inequality. In what follows we shall denote it SCG for short.

It is a three-setting inequality which is given in witness form by
\begin{equation*}
\begin{split}
SCG^{\rm eff} = & O^{\rm eff}_{A_1} \otimes ( \mathbb{1} + O^{\rm eff}_{B_1} + O^{\rm eff}_{B_2} + O^{\rm eff}_{B_3}) \\
+ & O^{\rm eff}_{A_2} \otimes ( \mathbb{1} + O^{\rm eff}_{B_1} + O^{\rm eff}_{B_2} - O^{\rm eff}_{B_3}) \\
+ & O^{\rm eff}_{A_3} \otimes (O^{\rm eff}_{B_1} - O^{\rm eff}_{B_2}) \\
+ & \mathbb{1} \otimes (O^{\rm eff}_{B_1} + O^{\rm eff}_{B_2}),
\end{split}
\end{equation*}
which must obey, for an initial two-particle state $ \rho $
\begin{equation}
Tr(SCG^{\rm eff}\rho) \geq -4.
\end{equation}
We tested for various measurement settings and the results can be found in Table II, in Appendix B.

It is interesting to look at the dependence of the lifetime of the violation
(in units of 10 ns)
in terms of $\epsilon$ (Fig.~\ref{fig2}) and the maximal violation in terms of
$\epsilon$ (Fig.~\ref{fig3}). For the CHSH, Alice measures her fixed quasi-spin $\Ket{\bar{K}^0}$ at time $t_{A1} = \tau, t_{A2} = 0 $ (in fact all measurements assume a fixed $\Ket{\bar{K}^0}$ quasi-spin, modifying only measurement times, for both parties)
and Bob measures at $t_{B1} = 0, t_{B2} = \tau $. For the SCG, Alice measures at $t_{A1} = 0, t_{A2} = \tau ,  t_{A3} = 2\tau $ and Bob measures at $t_{B1} = 0, t_{B2} = 2\tau ,  t_{B3} = \tau $.
Here, $\tau$ is just a plot parameter.
Plots of the maximal and minimal eigenvalues of the effective operator, as functions of time, for different measurement settings, are attached in Table I in Appendix B.

For $ \epsilon = 0 $ the maximal values of the effective operators were $ S^{\rm eff} \approx 2.12$ and $ SCG^{\rm eff} \approx -5.58 $. For $ \epsilon = 10^{-3} $ the value for the CHSH was $ S^{\rm eff} \approx 2.11$. However, the SGC inequality showed a remarkable robustness to the variation of the CP-violation parameter, for values of $\epsilon = 0.2$ one can still observe an effective operator eigenvalue of $ SCG^{\rm eff} \approx -4.49 $ (see Table II in Appendix B).

We notice that, while the presence of CP-violation does reduce the amount of violation we observe, while testing Bell-type inequalities, the violation is still present for values of $ \epsilon $ much larger than the one characteristic for neutral kaons.

\section{Simulating kaon-like behaviour with trapped Ions}

\subsection{General requirements }

\begin{figure}
  \centering
    \includegraphics[width=0.40\textwidth]{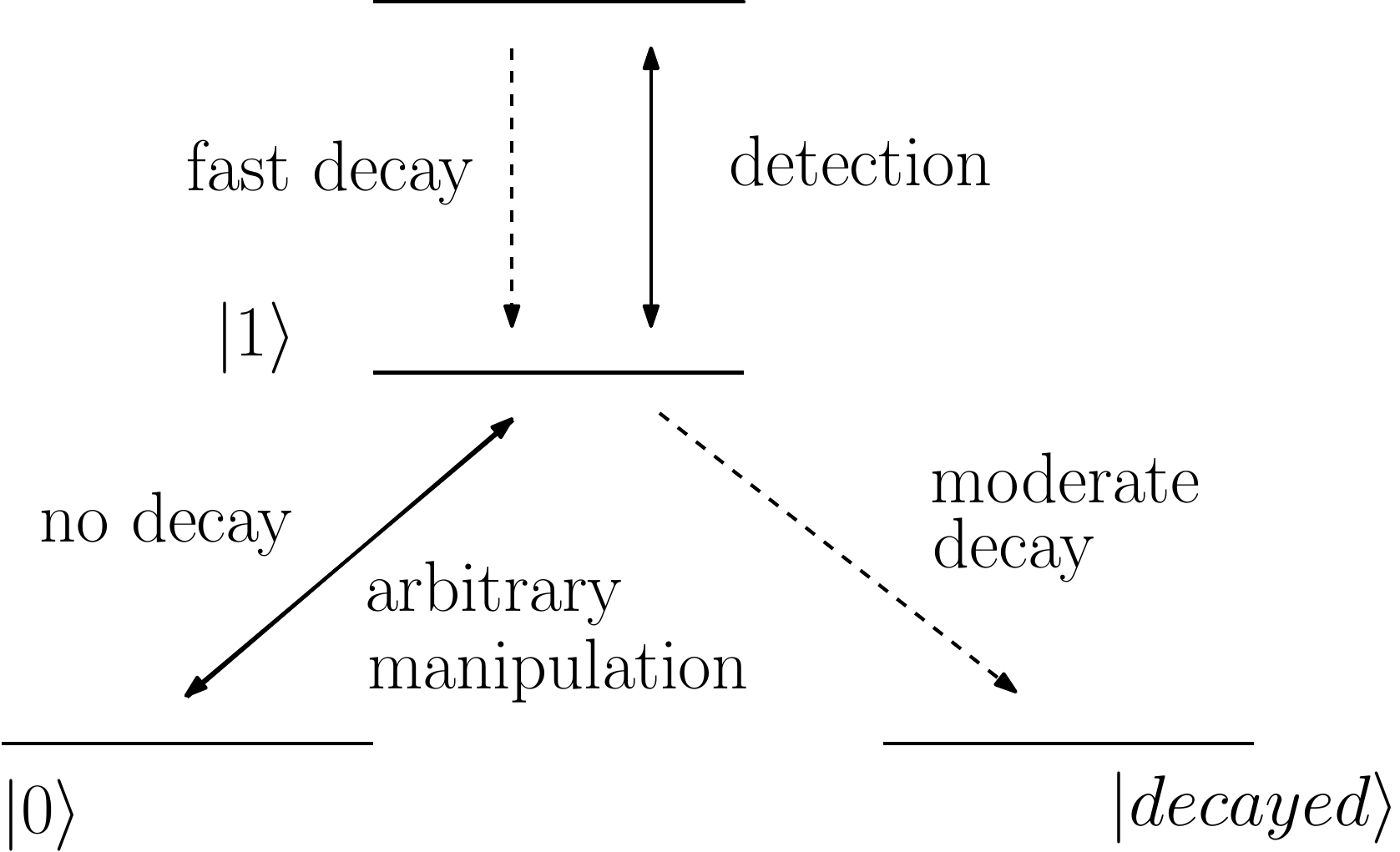}
    \caption{Basic scheme needed to implement kaon-like behaviour}
  \label{fig:fig4}
\end{figure}

\par Before looking at two examples of kaon-like behaviour in trapped Yb ions we summarize the general requirements for simulating kaon-like behaviour with ions. 

As shown in Fig.~\ref{fig:fig4}, there must be two levels which we identify as $\Ket{1}=\Ket{K_1^0}$ and $\Ket{0}=\Ket{K_2^0}$ and consecutively $\Ket{K_0}=\Ket{+}$ and $\Ket{\bar{K}_0}=\Ket{-}$ with $\Ket{\pm}= (\Ket{0}\pm\Ket{1})/\sqrt{2}$. There should be no decay between these two levels (as there is no decay between $\Ket{K_1^0}\approx \Ket{K_S}$ and $\Ket{K_2^0}\approx\Ket{K_L}$). Beside a two-qubit gate to create entanglement also arbitrary singe-qubit rotations (for example by using an RF-field) are needed for preparation, inducing kaon oscillation and choosing arbitrary measurement directions. 

Without CP violation, the oscillation between $\Ket{K^0}$ and $\Ket{\bar{K}^0}$ corresponds to an oscillation around the $z-$axis given by $U(t)=\exp(-i \delta t \sigma_z)$ with the Pauli matrix $\sigma_z$. The rotation frequency $\delta$ is given by the detuning $\delta=\omega_K-\omega_L$ between the level splitting $\omega_K$ and the reference laser $\omega_L$ determining the rotating reference frame.

The CP violation $\varepsilon$ leads to several effects. For example $\Ket{K^0(t)}$ is decaying faster than $\Ket{\bar{K}^0(t)}$ and the decay rate for both states start to oscillate. Furthermore their expectation values $\langle \sigma_z(t)\rangle$ undergo small oscillations and we find $\langle \sigma_z(t)\rangle_{K^0}\geq \langle \sigma_z(t)\rangle_{\bar{K}^0}$ for all times. This behaviour can be simulated by slightly tilting the rotation axis. The connection between the CP violation and the tilting is given by
\begin{equation}
\frac{\delta}{\sqrt{\delta^2+\Omega^2}}=\frac{1-\varepsilon}{1+\varepsilon}
\end{equation}
with the Rabi frequency $\Omega$ determining the strength of the laser or microwave.

 Two additional levels are needed, one for fluorescence detection (from which a fast decay must exist to one of the qubit states) and another, representing the state of decay products, with a moderate decay rate $\Gamma_S$ from one of the two qubit states.

In general, arbitrary values for the oscillation frequency $\omega$, the decay rate $\Gamma_S$ and the CP-violation $\varepsilon$ can be choosen for our ion system. However, the behaviour of the here described ion system mimics kaon-like behaviour only for small values of the CP violation, that is $\varepsilon \ll \omega/\Gamma_S$. This becomes apparent if we look at Eq.~(\ref{eq:kskl}) and Eq.~(\ref{eq:10}) which lead to unphysical behavour for large $\varepsilon$.

\subsection{\label{sec:level2} Examples}

\begin{figure}
  \centering
    \includegraphics[width=0.40\textwidth]{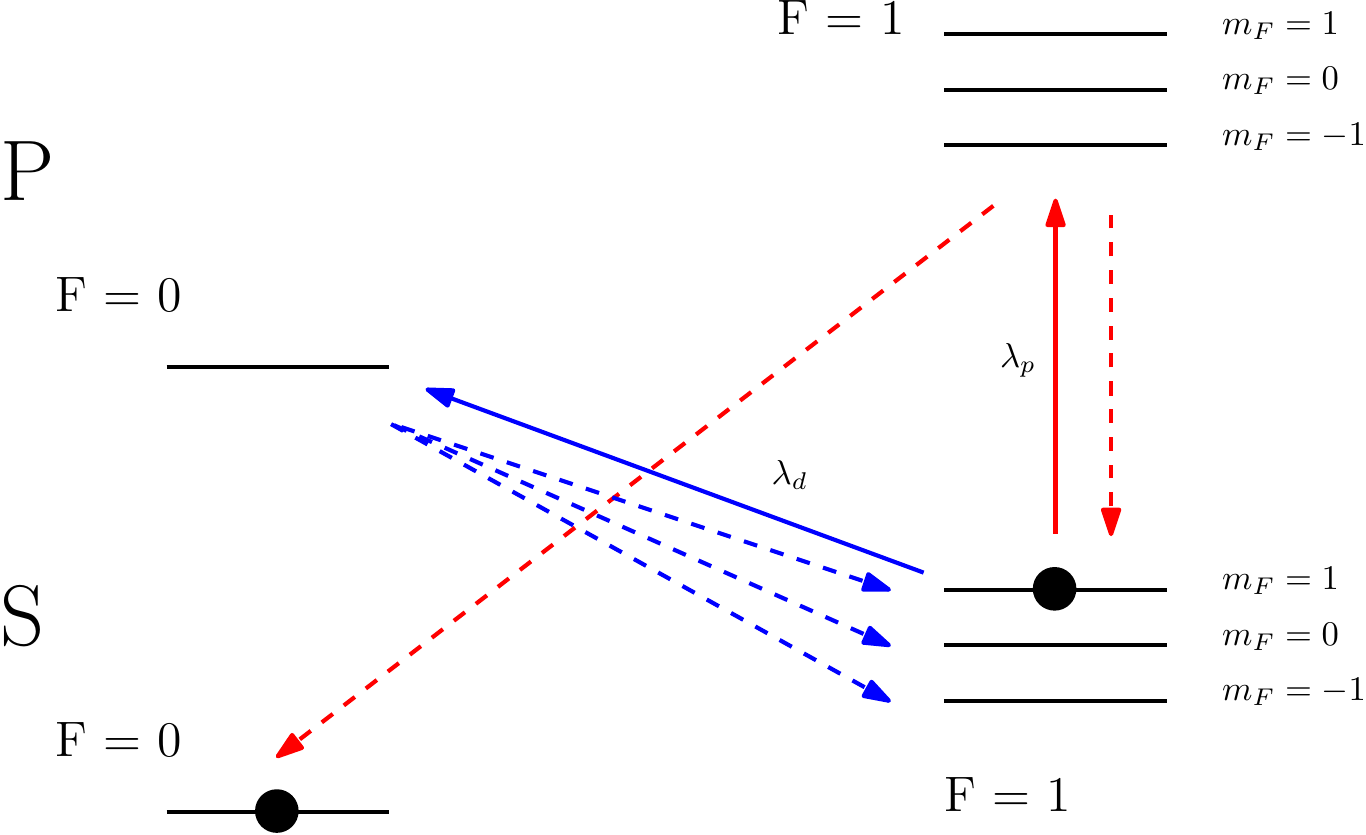}
    \caption{An implementation of the effective formalism using $^{171}Yb^+$. The qubit is defined between the two levels marked with thick dots.}
  \label{fig:ex1}
\end{figure}

\begin{figure}
  \centering
    \includegraphics[width=0.45\textwidth]{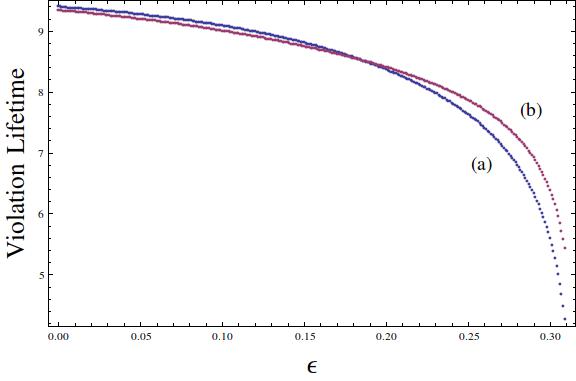}
    \caption{The effect of dephasing on the violation lifetimes (units of 10ns) for the SCG inequality, in terms of $\epsilon$. Here we plot both the case of (a) only decay with strength $\gamma_S$  and (b) splitting $\gamma_S$ into $2/3$ decay and $1/3$ dephasing.}
  \label{fig:lifetimecomp}
\end{figure}

\par In this final section we present two possible ways to simulate decaying systems with quasi-kaon behaviour using trapped  Yb ions. We will only briefly go over the examples here. For a more detailed description see the Appendix.

\par \textbf{Example 1: } The first proposal is based on the level structure of $^{171}Yb^+$  sketched in Fig.~\ref{fig:ex1}. The qubit is defined as $\Ket{0} = \Ket{S,F=0}$ and $\Ket{1} = \Ket{S,F=1, m_F = 1}$ and the decayed state is represented by $\Ket{S, F=1, m_F = 0}$ and $\Ket{S, F=1, m_F = -1}$.
 $\Ket{0}$ and $\Ket{1}$ are both long-lived states. However, decay can be generated by weak  driving of the $\Ket{S, F=1, m_F = 1} \leftrightarrow	 \Ket{P, F=0}$ transition with $\sigma^-$ polarized light. From  $\Ket{P, F=0}$ the state decays fast to all $\Ket{S, F=1,m_F}$ states. It is important to note here that the strength of the decay is thus tunable, by tuning the strength / duration of the transitions.

\par This decay behaviour is slightly different from kaons, because $\Ket{0}$ does not decay and there is a non-zero probability of $\Ket{1}$  "decaying" to itself, which leads to dephasing. More clearly, while for neutral kaons one could express the decay process as
\begin{equation}\label{eq:35}
\Gamma = \gamma_S \Ket{decayed}\Bra{1} + \gamma_L \Ket{decayed}\Bra{0},
\end{equation}
where, for clarity of notation, the short-lived kaon state has been identified with $\Ket{1}$ and long-lived kaon state with $\Ket{0}$; the case for the example 1 is described by
\begin{equation}\label{eq:36}
\Gamma' = \gamma_S \left( \frac{2}{3} \Ket{decayed}\Bra{1} +  \frac{1}{3}\Ket{1}\Bra{1}\right),
\end{equation}
where there is always a chance of decay to the initial level (dephasing). In Fig. 7 we plot the violation lifetimes in terms of the CP-violation parameter with and without splitting $\gamma_S$ into $2/3$ decay and $1/3$ dephasing.

Another necessary step to simulate kaon is to generate entanglement between two ions. This is achieved in our example by using MAGIC (Magnetic Gradient-Induced Coupling) \cite{magic1,magic2}.

State detection of the ion is achieved by a single qubit rotation to choose the measurement direction and consecutively driving the $\Ket{S,F=1} \leftrightarrow \Ket{P,F=0}$ transition with unpolarized light and detecting the scattered photons.Unfortunately, this state dependent fluorescence measurement is only able to distinguish between the states $\Ket{S,F=0}$ and $\Ket{S,F=1}$, but cannot resolve the sublevels $\Ket{m_F=0,0/\pm 1}$. Therefore, each probability measured in such a way will correspond to the sum of the probability $P(\Ket{k_n})$ to be in the state $\Ket{k_n}$ plus the probability that the ion/kaon has already decayed. Therefore, instead of measuring $P(\Ket{\bar{K}_0})$ we rotate $\Ket{K_0}=\Ket{+}$ onto the state $\Ket{1}$ and perform  in this way an inverse measurement and determine the propbability $P=1-P(\Ket{\bar{K}_0})$. That is, instead of asking "What is the probability corresponding to the state $\Ket{\bar{K^0}}$, we may equivalently ask "What is the probability for not obtaining $\Ket{\bar{K}_0}$?".

\par
\begin{figure}
  \centering
    \includegraphics[width=0.45\textwidth]{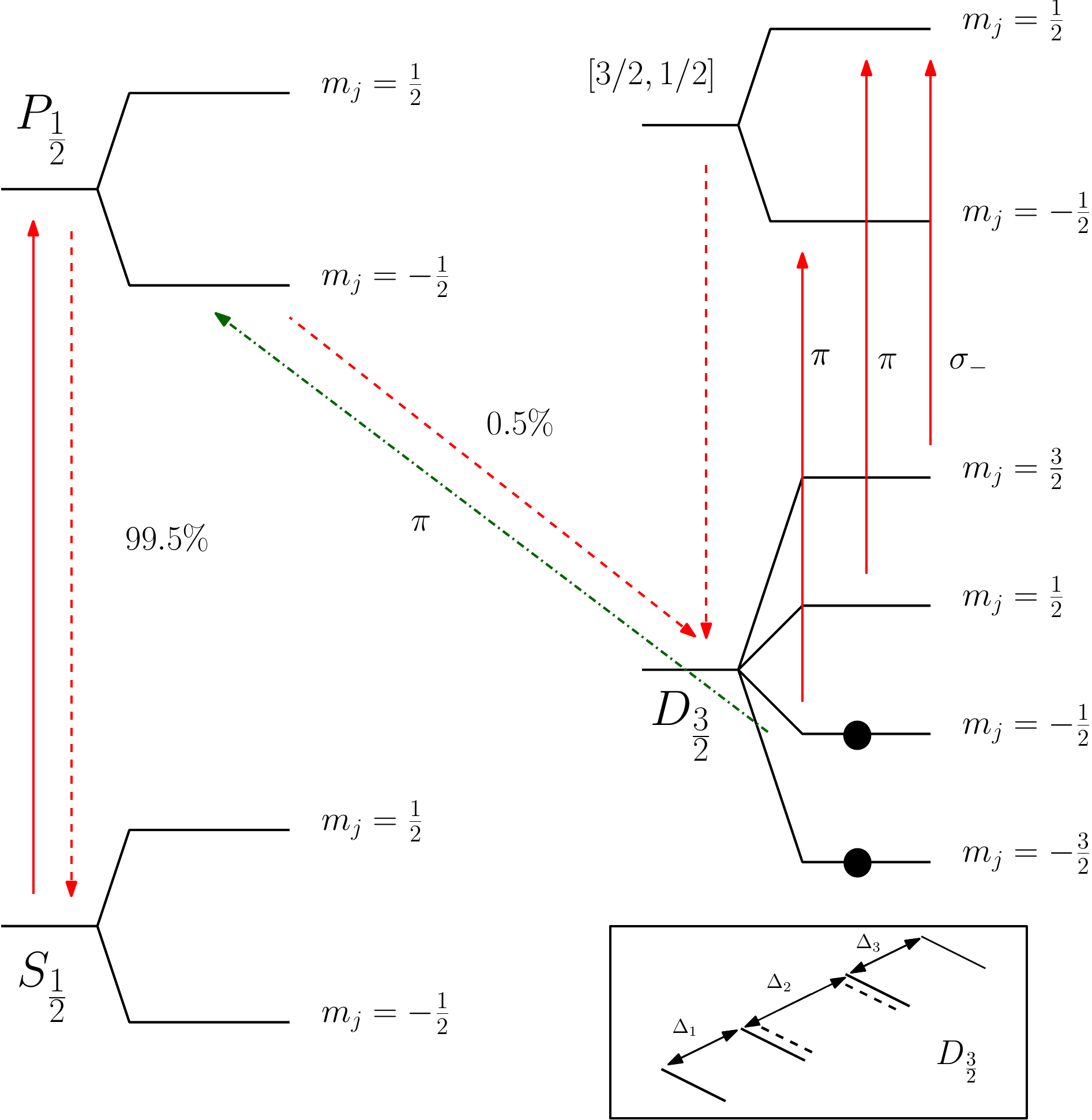}
    \caption{An implementation of the effective formalism using  $^{172}$Yb$^+$. The qubit is defined between the two levels marked with thick dots. The effect of the AC Stark shift is also shown in the lower box.}
  \label{fig:ex2}
\end{figure}

 \textbf{Example 2: } The second example comes to eliminate the dephasing problems one finds with the first. This time, we use $^{172}Yb^+$ ions (see Fig.~\ref{fig:ex2}). The qubit is defined as $\Ket{0}=\Ket{D_{3/2},m_j=-3/2}$ and $\Ket{1}=\Ket{D_{3/2},m_j=-1/2}$.
\par To simulate decay, we drive the $\Ket{D_{3/2}, m_j = - 1/2} \leftrightarrow \Ket{P, m_j = -1/2}$ transition with $\pi$-polarized light. From $\Ket{P, m_j = -1/2}$ the ion decays in over $99\%$ into the $\Ket{S,m_j=\pm 1/2}$ state and not back into the $\Ket{D_{3/2}, m_j = - 1/2}$Similar to the first example, we perform again an inverse measurement by transferring $\Ket{K^0}$ onto $\Ket{P, m_j = -1/2}$.

\par Two remarks need to be made here. First, the time evolution of "decay" and "oscillation" commute only for $\varepsilon=0 $. In this case, we are able to switch on the lasers/microwaves causing the decay and the oscillation one after the other. However, for $\varepsilon \neq 0$ our way to model interfere with the oscillation. Therefore, we have to use the Trotter theorem and approximated the time evolution by switching between oscillation and decay in short time intervals. This is a standard method in digital quantum simulations and the approximation can be made arbitrarily good by shorting e.g. the time intervals (see e.g. \cite{digisim}).

Second, as explained above, there are different decay channels between kaons and the chosen ion examples [compare Eq.(\ref{eq:35}) and Eq.(\ref{eq:36})]. This does not significantly modify the nature results for small CP violations $\epsilon$ as depicted in Fig.\ref{fig:lifetimecomp}.

\section{Conclusion}

We have described an easy-to-use formalism that facilitates the study of
entanglement for systems under non-unitary time evolution. Besides decay,
the systems chosen also display an oscillation between two orthogonal
states, raising the interesting possibility of performing the bipartite
measurements at different times. The generalized formalism was applied
to neutral kaons, in order to show that it does indeed reproduce previous
results, and then applied to the case of two Ytterbium isotopes, $^{171}$Yb
and $^{172}$Yb. The purpose of the latter is to exemplify a similar type
of behaviour, which comes naturally for kaons, on a different, but practically
relevant system; the motivation given by the fact that trapped ions are an important
implementation for quantum computation in particular and quantum information
processing in general.

The treatment of entanglement in unstable systems can provide a new way of studying
the predictions of quantum mechanics in systems other than kaons and ions. The
behaviour treated above can be reproduced with other systems, for example
photons travelling through optical fibres. In this case birefringence
determines fast and slow polarization modes, an analogue of the long
and short-lived states of neutral kaons, and polarization dependent
loss is an analogue to the decay property \cite{go}.

Meson-antimeson systems exhibit one interesting feature that
does affect the amount of violation one observes in Bell-type
inequalities, the phenomenon of CP-violation. Because CP-violation
translates into an asymmetry between matter and antimatter the
eigenstates of the system's Hamiltonian become slightly non-orthogonal
(due to the different probabilities corresponding to the states
representing the particle and its antiparticle). Such behaviour
can also be simulated in the case of trapped ions by an imperfect
preparation of the initial Bell-state.
Another essential feature of the formalism is that it allows
for an analytical method to be applied (see Eq.~(\ref{eq:26})),
because it reduces the entire time evolution of the system to
exponentiating one single operator that encompasses both
unitary and non-unitary time evolution.

For future research, it would be desirable to use the formalism for other
systems where decay or no-detection events play a role, such as polarized
photons. Then, the formalism can be combined with other tools in entanglement
theory, such as entanglement witnesses. This may open a way for entanglement
characterization and quantification in the presence of noise and imperfect
detectors.

We thank Ali Asadian and Michael Johanning for the useful discussions and
suggestions. This work has been supported by
the FQXi Fund (Silicon Valley Community Foundation),
the DFG Research Unit FOR 1873, and the ERC (Consolidator Grant 683107/TempoQ).

\section*{APPENDIX A}

\par \textbf{Example 1: }The first example (Fig.~\ref{fig:ex1}) uses $^{171}$Yb$^+$ ions. The qubit is defined here between $\Ket{S, F=0} = \Ket{0}$ and $\Ket{S,F=1,m_F = 1} = \Ket{1}$ (highlighted in the figure by thick dots).
\par The initialization is done by driving an unpolarized pumping laser, $\lambda_p$ between the $F=1$ levels, of $S$ and $P$, and a consecutive decay to the $\Ket{S, F=0}$ state. The arbitrary rotation in  the qubit basis $\{\Ket{0},\Ket{1}\}$ is performed via a polarized RF-field. In this way, any superposition of $\Ket{0}$ and $\Ket{1}$ can be prepared. This also assures that a readout is possible in any basis (corresponding to a rotation from the $\{K_S, K_L\}$ to the $\{K^0, \bar{K^0}\}$, in the kaon case).
\par The decay is simulated by a second polarized laser $\lambda_d$ driving the transition $\Ket{S, F=1, m_F = 1} \leftrightarrow	 \Ket{P, F=0}$. From $\Ket{P, F=0}$, the ion decays very fast into the  $\Ket{S, F=1}$ levels, the probabilities for each of the $m_F = 0,\pm 1$ are equal.  This essentially turns the $\Ket{F=1, m_F = 0}$ and $\Ket{F=1, m_F = -1}$ sublevels into a generalized decayed state (the third state in our kaon formalism). There is an inconvenience here, however. Due to the equal probability of a decay from the $P$ level back to any of the three $\Ket{S, F=1}$ sublevels, one also gets dephasing. We will see in the second example how this problem can be overcome.
\par The essence of the effective formalism is that it does not distinguish between a non-detection event and one of the two possible states (decayed or $\Ket{K^0}$ , when measuring $\Ket{\bar{K^0}}$, in the kaon case). This could be translated into
\begin{equation}
\underbrace{P_{\Ket{\bar{K^0}}}}_{\text{P(Y)}} + \underbrace{P_{\Ket{K^0}} + P_{decayed}}_{\text{P(N)}} = 1.
\end{equation}
\par This provides the option of performing the opposite measurement (corresponding to $P(N)$ above). For this , we perform a population inversion between the $\Ket{S, F=1}$ sublevels and the $\Ket{S, F=0}$ level. Then, a typical fluorescence measurement can be performed on the $S$ level.
There is also a second option to perform the population inversion and that is a population shelving from the two sublevels, representing the decayed state, to some other atomic level.

\par \textbf{Example 2: }A second way to implement the formalism is to use $^{172}Yb^+$ ions. In this case, the qubit is defined between the $\Ket{D_{3/2},m_j=-3/2}$ and $\Ket{D_{3/2},m_j=-1/2}$ levels (Fig.~\ref{fig:ex2}).
\par The implementation is done by running a pump laser between the $S$ and $P$ states, this leads to a population transfer to all four $D_{3/2}$ sublevels, due to decay. Then a combination of $\pi$ and $\sigma_-$ polarized lasers move the populations of the upper three $D$ sublevels to the $\Ket{[3/2,1/2]}$ states, and ultimately to the $\Ket{D_{3/2},m_j=-3/2}$ sublevel.
\par Coherent driving of the $\Ket{D_{3/2},m_j = -3/2} \leftrightarrow \Ket{D_{3/2},m_j = -1/2} $ transition causes a problem, because the level splitting of all $\Ket{D_{3/2}}$ states are equal. To isolate this transition, an AC Stark shift is induced, to increase the distance between the qubit levels and the sublevels $m_j = 1/2$ and $m_j = 3/2$.
\par The decay is modelled by a weak pulse driving the transitions $\Ket{D_{3/2},m_j=-1/2} \leftrightarrow \Ket{P_{1/2}}$.
\par Finally, we transfer again the state $\Ket{K^0}$ onto $\Ket{P}$ before performing a state dependent fluorescence measurement by driving the $\Ket{S} \leftrightarrow \Ket{P}$ transition. This measurement is again an inverse measurement similar to the first example, essentially measuring the probability $1 - P(\bar{K}^0)$ (using kaon notation).
\par In  both examples, two ions can be entangled with the help of MAGIC \cite{magic1,magic2} to generate a bipartite system with similar properties as pairs of neutral kaons.


\begin{table*}[ht]\label{tab1}
  \centering
    \textbf{\normalsize{APPENDIX B}}\par\medskip
  \vspace{1cm}
\hspace{-0.5cm}
\begin{tabular}{|c|c|c|}
\hline
$\epsilon$  & Alice (0,t), Bob(t,0) & Alice(t,0),Bob(0,t)\tabularnewline
\hline
0 &\includegraphics[width=0.47\textwidth]{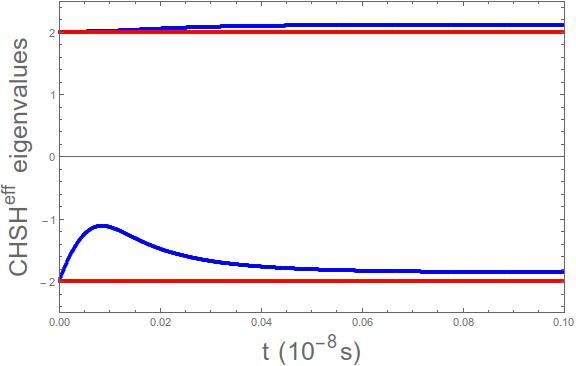}  &\includegraphics[width=0.47\textwidth]{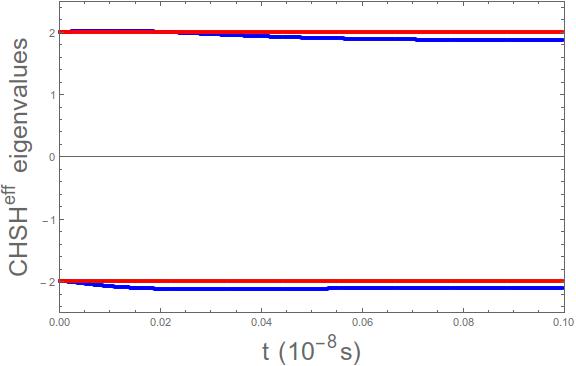} \tabularnewline
\hline
$10^{-3}$ & \includegraphics[width=0.47\textwidth]{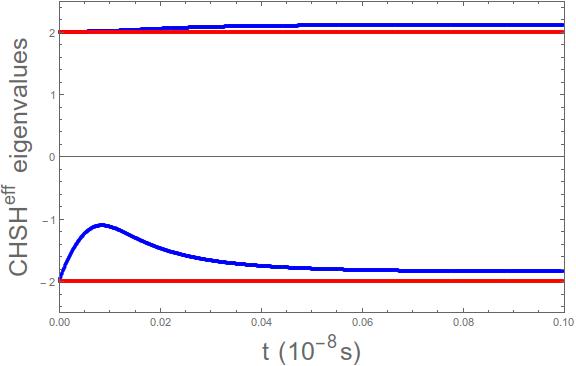}  & \includegraphics[width=0.47\textwidth]{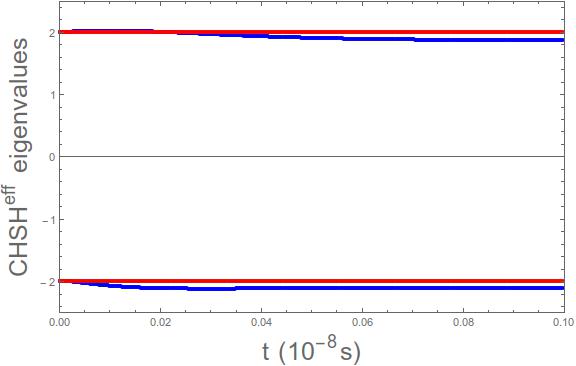}\tabularnewline
\hline
$10^{-1}$ & \includegraphics[width=0.47\textwidth]{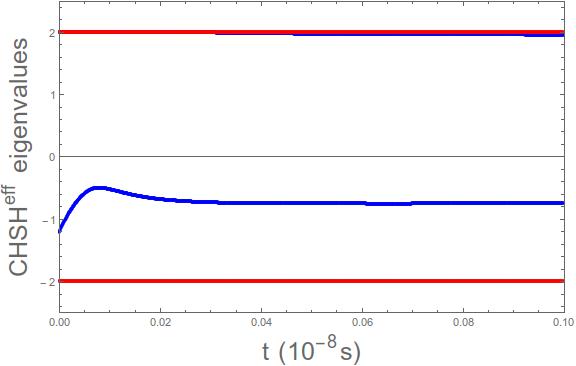} & \includegraphics[width=0.47\textwidth]{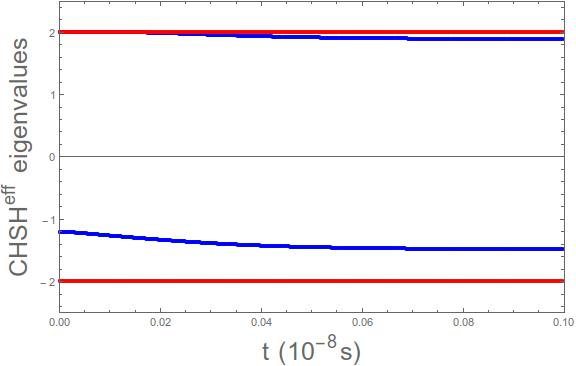} \tabularnewline
\hline
\end{tabular}
\caption{ Plots (for neutral kaons) for the minimal and maximal eigenvalues of the CHSH effective operator (represented here in blue) for certain values of the CP-violation parameter. The red lines represent the classical limits of the CHSH inequality. The settings are for both Alice and Bob measuring at different times, for example A(0,t) means Alice's first measurement is made at $\tau = 0$ and the second measurement at $\tau =t $, where t is a plot parameter.}
\end{table*}

\begin{table*}[ht]\label{tab2}
\hspace{-0.5cm}
\begin{tabular}{|c|c|c|}
\hline
$\epsilon$  & Alice (0,t,t), Bob(t,t,0) & Alice(0,t,2t),Bob(0,2t,t)\tabularnewline
\hline
0 &\includegraphics[width=0.47\textwidth]{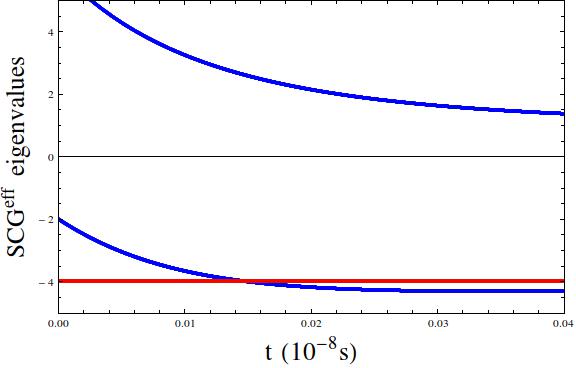}  &\includegraphics[width=0.47\textwidth]{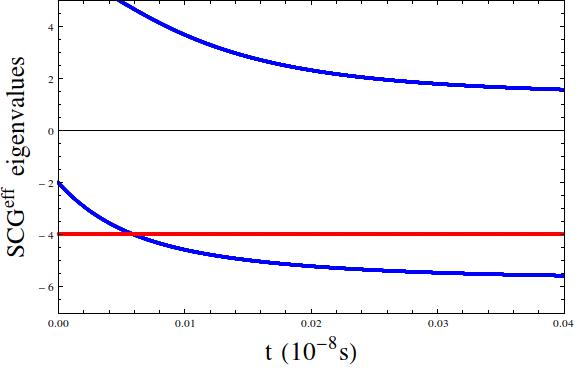} \tabularnewline
\hline
$10^{-2}$ & \includegraphics[width=0.47\textwidth]{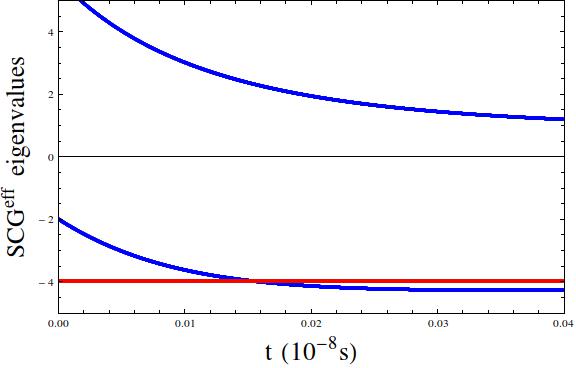}  & \includegraphics[width=0.47\textwidth]{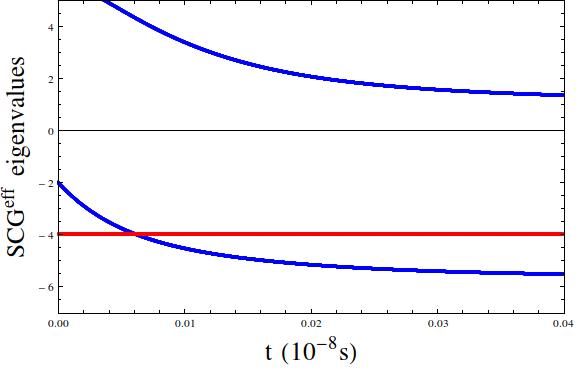}\tabularnewline
\hline
$0.2$ & \includegraphics[width=0.47\textwidth]{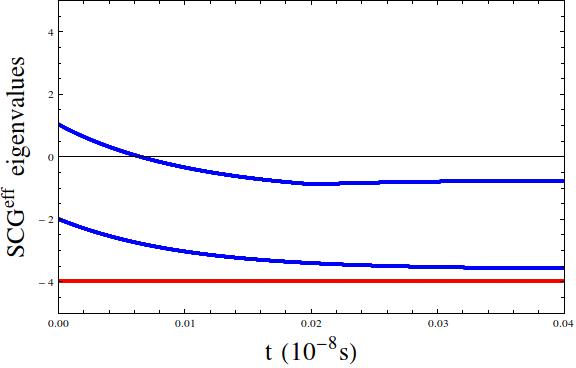} & \includegraphics[width=0.47\textwidth]{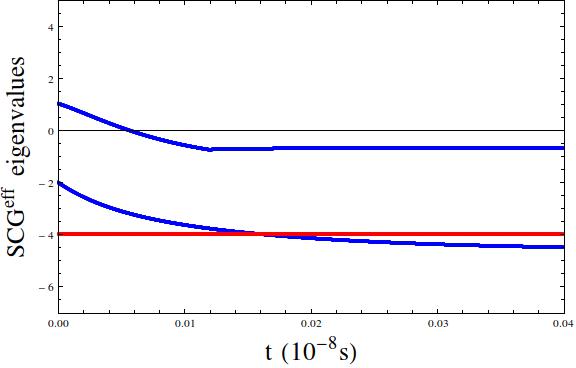} \tabularnewline
\hline
\end{tabular}
\caption{ Plots (for neutral kaons) for the minimal and maximal eigenvalues of the SCG effective operator (represented here in blue) for certain values of the CP-violation parameter. The red line represents the classical limit of the SCG inequality.}
\end{table*}

\end{document}